\documentclass[aps,reprint,superscriptaddress]{revtex4-2}
\usepackage{graphics}
\usepackage{hyperref}
\usepackage{mathtools}

\newcommand{\ndP}{\widetilde{P}}
\newcommand{\ndOME}{\widetilde{\Omega}}
\newcommand{\ndome}{\widetilde{\omega}}
\newcommand{\ndt}{\tilde{t}}
\newcommand{\ndr}{\tilde{r}}
\newcommand{\ndRin}{\widetilde{R}_{\rm in}}
\newcommand{\ndRout}{\widetilde{R}_{\rm out}}
\newcommand{\ndv}{\tilde{v}}

\begin{document}
\title{Mutual information as a measure of mixing efficiency in viscous fluids}

\author{Yihong Shi}
\affiliation{Max Planck Institute for Dynamics and Self-Organization (MPI-DS), Am Fassberg 17, 37077 G\"{o}ttingen, Germany}

\author{Ramin Golestanian}
\email{ramin.golestanian@ds.mpg.de}
\affiliation{Max Planck Institute for Dynamics and Self-Organization (MPI-DS), Am Fassberg 17, 37077 G\"{o}ttingen, Germany}
\affiliation{Rudolf Peierls Centre for Theoretical Physics, University of Oxford, Oxford OX1 3PU, UK}
\affiliation{Institute for the Dynamics of Complex Systems, University of G\"{o}ttingen, 37077 G\"{o}ttingen, Germany}

\author{Andrej Vilfan}
\email{andrej.vilfan@ds.mpg.de}
\affiliation{Max Planck Institute for Dynamics and Self-Organization (MPI-DS), Am Fassberg 17, 37077 G\"{o}ttingen, Germany}
\affiliation{Jo\v{z}ef Stefan Institute, 1000 Ljubljana, Slovenia}

\date{\today}

\begin{abstract}
Because of the kinematic reversibility of the Stokes equation, fluid mixing at the microscale requires an interplay between advection and diffusion. Here we introduce mutual information between particle positions before and after mixing as a measure of mixing efficiency. We demonstrate its application in a Couette flow in an annulus and show that non-uniform rotation sequences can lead to more efficient mixing. We also determine mutual information from Brownian dynamics simulations using data compression algorithms. Our results show that mutual information provides a universal and assumption-free measure of mixing efficiency in microscale flows.
\end{abstract}

\maketitle
Designing protocols that force an out of equilibrium system into equilibrium faster than the natural relaxation rate is a pertinent topic in a number of classical \cite{Martinez.Ciliberto2016,Lu2017} and quantum systems \cite{Berry2009}. 
A prime example of forced equilibration is the mixing of fluids if the initial state contains a nonequilibirum concentration or temperature distribution. Fluid mixing at the microscale is of paramount importance in biological organisms and in artificial systems. Examples range from the uptake of oxygen, nutrients or chemical signals in aquatic organisms to microreactors and ``lab on a chip'' applications \cite{Stroock.Whitesides2002,Ottino.Grzybowski2004, Grigoriev.Sharma2006,Pine.Leshansky2005,Aref.Tuval2017}. In biology, mixing is frequently accomplished by cilia which drive long-range flows, but also localized regions of chaotic advection \cite{Supatto.Vermot2008,Uchida2010,Rahbar.Gray2014,Ding.Kanso2014, Jonas.Choma2013, Nawroth.McFallNgai2017}. A particular challenge to microscale mixing is posed by the time-reversibility of flows at low Reynolds numbers \cite{Golestanian2011,Arrieta.Tuval2020}. Mixing therefore requires an interplay between advection (stirring) and diffusion \cite{Villermaux2019,Tang.Golestanian2020}. Although most examples work with the mixing of two distinct fluids, the formalisms that are used apply equally to other scalar quantities like temperature \cite{Villermaux2019}.

The measures that have been proposed to quantify the mixing efficiency can broadly be classified as global and local \cite{Aref.Tuval2017,Villermaux2019}. Global measures typically start by imposing a pattern, e.g., by distributing the solute in a part of the fluid volume. After mixing, the distribution of the solute can be measured by defining the intensity of segregation as the variance of solute concentration (the $L^2$ norm) \cite{Danckwerts1952,Thiffeault2012}, its entropy \cite{DAlessandro.Mezic1999, Camesasca.ManasZloczower2006,Brandani.Aegerter2013,Kree.Zippelius2019}, the mean distance to the closest particle from the other population \cite{Stone.Stone2005} or Sobolev norms \cite{Mathew.Petzold2005,Thiffeault2012}.
The same global mixing measures can also be applied to quantify unmixing in cases of spontaneous phase segregation \cite{Brandani.Aegerter2013}. The limitation of these measures is that the result will depend on the choice of the initial distribution.
Local measures typically characterize the amount of stretching or the Lyapunov exponents \cite{Tsang.Ott2005,Arrieta.Tuval2020,Meunier.Villermaux2022}.

\begin{figure}						
\includegraphics[width =\columnwidth]{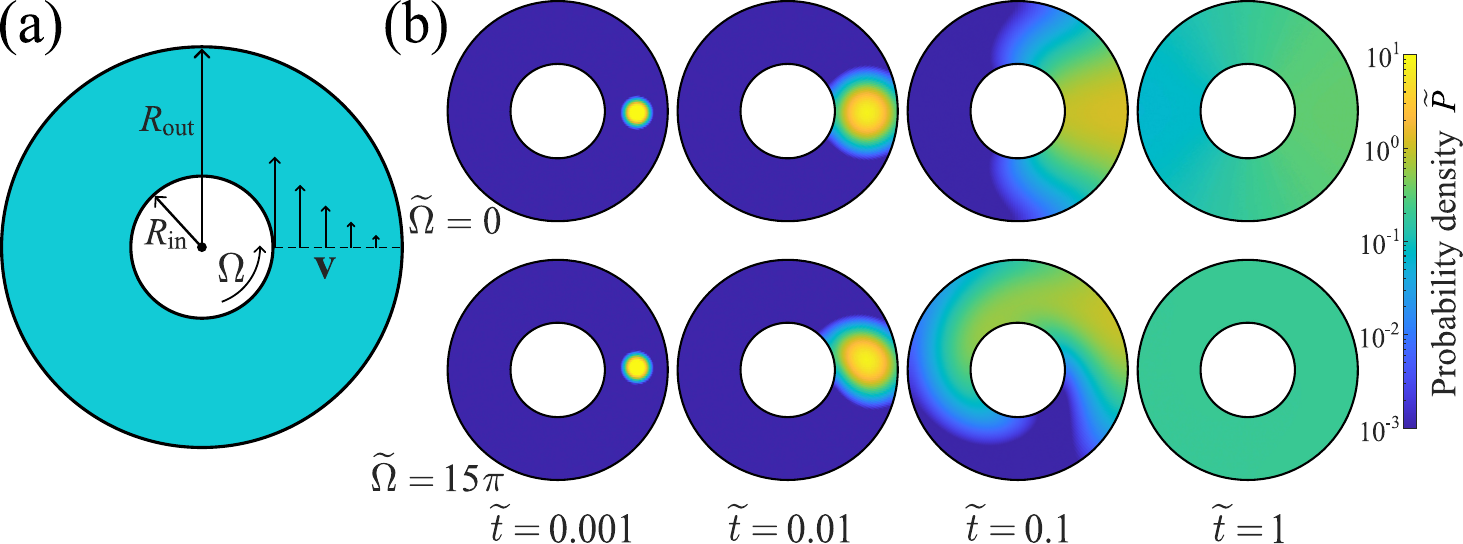}
\caption{(a) 2D Couette flow in an annulus geometry. The outer boundary is stationary and the inner is rotating with angular velocity $\Omega$. (b) Time evolution of the probability density $\ndP(\mathbf{x},\ndt\vert \mathbf{x}_0)$, with an initial position $\mathbf{x}_0$ in the middle of the annulus. The top row shows the process with pure diffusion ($\ndOME=0$) and the bottom row  with both diffusion and uniform advection ($\ndOME=15\pi$).}
\label{fig1}		
\end{figure}

In this paper, we introduce mutual information as a universal measure of mixing in fluids at low Reynolds numbers with strong interplay between advection and diffusion. As a simple model system, we test our method in a 2D Couette flow in an annulus geometry and show that the mixing efficiency depends in a non-trivial way on the time sequence of rotation (see Fig.~\ref{fig1}).

Because the theoretically smallest compressed size of a dataset is given by its Shannon entropy, lossless data compression algorithms can be used to estimate the entropy of a distribution \cite{Schurmann.Grassberger1996,Avinery.Beck2019,Martiniani.Levine2019}. Order in two-dimensional systems can be estimated with lossless image compression algorithms like PNG \cite{Ziepke.Frey2022} or GIF with LZW compression \cite{Ariel.Diamant2020}. However, it has also been argued that compression algorithms lead to poor estimates in two-dimensional systems with long range correlations \cite{Brigatti.deSousaFilho2022,deSousaFilho.Brigatti2022}.
We estimate mutual information with different data compression algorithms and show that the neural-network based PAQ8PX \cite{paq8px} gives good accuracy.

The motion of the fluid at low Reynolds number is governed by the Stokes equation $\mu \Delta \mathbf{v} = \boldsymbol{\nabla} p$ and incompressibility condition $\boldsymbol{\nabla} \cdot \mathbf{v}=0$. Here, $\mathbf{v}$ is the fluid velocity, $p$ is the pressure and $\mu$ is the fluid viscosity. We assume that the shape of the fluid volume does not change during the mixing process, although an extension to shape-changing compartments is straightforward. Therefore, the normal component of the fluid velocity has to vanish at the boundary, $\mathbf{n}\cdot\mathbf{v}=0$. Because there are no inertial terms in the Stokes equation, the fluid motion at any time $t$ is fully determined by the instantaneous (tangential) velocities at the boundaries at the same time. The particles in the fluid to be mixed -- we assume that they are all equivalent -- are subject to Brownian motion in addition to the advection caused by the fluid flow. The time evolution of the probability density $P(\mathbf{x},t)$ to find the particle at position $\mathbf{x}$ at time $t$ is determined by the advection-diffusion equation (equivalent to a Fokker Planck equation)
\begin{equation}
     \label{1E}
     \partial_{t} P + \boldsymbol{\nabla}\cdot(\mathbf{v}P)=D\Delta P
\end{equation}
with the diffusion constant $D$.
The zero-flux condition further implies $\mathbf{n}\cdot \boldsymbol{\nabla} P(\mathbf{x},t)=0$ at the boundary. The conditional probability $P(\mathbf{x},t \vert \mathbf{x}_0)$ is obtained by solving Eq.~\eqref{1E} with the initial condition $\mathbf{x}=\mathbf{x}_0$ at $t=0$. Because the flow is divergence-free, the stationary solution is always given by a constant density, $P(\mathbf{x},t)\equiv 1/V$.

\newcommand{\finalstate}{I_{F}}

We quantify the efficiency of fluid mixing using the mutual information between the initial and the final position of a particle in the fluid. Mutual information provides the strictest possible measure of mixing: zero mutual information means that the final position of a particle is unrelated to its initial position and therefore also to the position of any other particle in the fluid. Unlike many other criteria, it does not require any assumptions about the initial spatial distribution of the fluid components to be mixed. Mutual information is defined as sum of the entropies of the distribution of the initial and final distributions of the position, reduced by the joint entropy of the initial and final position together \cite{infobook}:
\begin{equation}\label{2E}
    I[\mathbf{x}_{0};\mathbf{x}_{t}]=S[\mathbf{x}_{0}]+S[\mathbf{x}_{t}]-S[\mathbf{x}_{0}, \mathbf{x}_{t}].
\end{equation}
Here, $S[\mathbf{x}_{0}]$ is the entropy of the initial position variable, $S[\mathbf{x}_{t}]$ is the entropy of the position variable at time $t$, and $S[\mathbf{x}_{0}, \mathbf{x}_{t}]$ is the joint entropy of these two position variables.
The final distribution at time $t$ can also be expressed as $P(\mathbf{x},t)=\int P(\mathbf{x},t\vert \mathbf{x}_0) P(\mathbf{x}_0)d\mathbf{x}_0$.
Equivalent to Eq.~\eqref{2E}, the mutual information can be expressed in terms of the conditional entropy as
\begin{equation}\label{3E}
    I[\mathbf{x}_{0};\mathbf{x}_{t}]=S[\mathbf{x}_{t}]-S[\mathbf{x}_{t}|\mathbf{x}_{0}],
\end{equation}
where $S[\mathbf{x}_{t}] = -\int P(\mathbf{x},t)\log{P(\mathbf{x},t)} \,d\mathbf{x}$, and $S[\mathbf{x}_{t}|\mathbf{x}_{0}] = -\int{P(\mathbf{x}_{0})\left[\int{P(\mathbf{x},t|\mathbf{x}_{0})\log P(\mathbf{x},t|\mathbf{x}_{0})}d\mathbf{x}\right]}d\mathbf{x}_{0}$ is the conditional entropy with the knowledge of the initial position.

Mutual information, as we define it, depends on the distribution of initial positions $P(\mathbf{x}_{0})$. This distribution determines the statistical weight that is given to the mixing efficiency in different regions of the fluid. It should not be confused with any imposed pattern in the fluid to be ``erased'' by mixing, as frequently used in other mixing efficiency criteria. In the following, we assume a homogeneous weight $P(\mathbf{x}_{0})= 1/V$. As this is the stationary solution, it also leads to $P(\mathbf{x},t)= 1/V$ at all later times. We note, however, that as a possible alternative, one could also use an initial distribution that maximizes $I(t)$, following the spirit of Shannon's channel capacity theorem, which would provide a quantitatively stricter measure.

Interestingly, the mixing efficiency is invariant upon time-reversal of the mixing sequence, defined by $\bar{\mathbf{v}}(\mathbf{x},t)=-\mathbf{v}(\mathbf{x},t_F-t)$, where $t_F$ is the duration of the mixing process. This follows from a general property of the Fokker-Planck equation with divergence-free flux densities \cite{Dieball.Godec2022}, for which
    $\bar{P}(\mathbf{x}, t|\mathbf{x}_0)=P(\mathbf{x}_0, t|\mathbf{x})$.
With a uniform distribution $P(\mathbf{x}_0)$, we have $S[\mathbf{x}_t]=S[\mathbf{x}_0]$ and from Eq.~\eqref{3E} it follows immediately that $\bar{I}(t_F)=I(t_F)$.

We now demonstrate the application of mutual information as measure for mixing efficiency in 2D Couette flow. The outer circular boundary with radius $R_{\rm out}$ is stationary and the inner boundary with radius $R_{\rm in}$ rotates with an angular velocity $\Omega(t)$, as shown in Fig.~\ref{fig1}(a). In the following we introduce a characteristic length scale $L=(R_{\rm in}+R_{\rm out})/2$ and time scale $T=L^2/D$ and use them to non-dimensionalize the variables as
$\ndt=t/T$, $\ndr=r/L$, $\ndOME=\Omega T$, $\ndRin=R_{\rm in}/L$ and $\ndRout=R_{\rm out}/L$. 

In these variables, the 2D Stokes equation in the annulus is solved by $\ndv=\ndome(\ndr) \ndr$ with \cite{happel1983}
\begin{equation}\ndome(\ndr) = \ndOME \frac{\ndr^{-2}-\ndRout^{-2}}{\ndRin^{-2}-\ndRout^{-2}}\;.
\end{equation}

In order to numerically determine the conditional entropy  $S[\mathbf{x}_t|\mathbf{x}_0]$ in Eq.~\eqref{3E}, we need to calculate  $P(\mathbf{x},t|\mathbf{x_{0}})$ for any initial position $\mathbf{x}_0$. For this purpose, we express the advection-diffusion equation [Eq.~\eqref{1E}] in polar coordinates
\begin{equation}
     \label{6E}
     \frac{\partial \ndP}{\partial \ndt} = \frac{1}{\ndr}\frac{\partial \ndP}{\partial \ndr} + \frac{\partial^2 \ndP}{\partial {\ndr}^2} + \frac{1}{{\ndr}^2}\frac{\partial^2 \ndP}{\partial \theta^2} - \ndome(\ndr)\frac{\partial \ndP}{\partial \theta}
\end{equation}
along with the boundary condition $\partial\ndP/\partial\ndr=0$ at $\ndRin=0.6$ and $\ndRout=1.4$. We solve Eq.~\eqref{6E} numerically with a spectral method. Two examples of the solution for $P(\mathbf{x},t|\mathbf{x_{0}})$, one with pure diffusion and one with uniform rotation of the inner boundary, are shown in  Fig.~\ref{fig1}(b). 
The conditional entropy $S[\mathbf{x}_t|\mathbf{x}_0]$, needed to determine the mutual information $I(t)$, is eventually obtained by integrating $P(\mathbf{x},t|\mathbf{x_{0}}) \log P(\mathbf{x},t|\mathbf{x_{0}})$ over $\ndr_0$, $\ndr$ and $\theta$ and making use of the rotational symmetry of the system with regard to $\theta_0$.

\begin{figure}
\includegraphics[width=0.8\columnwidth]{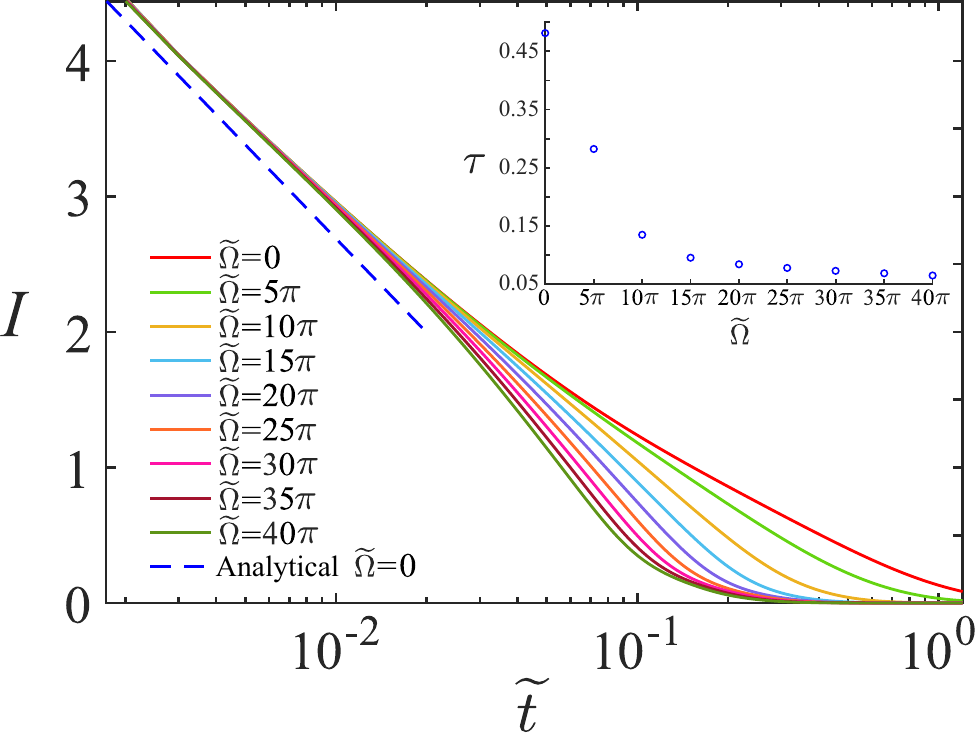}
\caption{Temporal decay of mutual information between particle positions in the initial state and the state after time $\ndt$ for different rotation rates. The dashed line shows the approximation for diffusive motion, $I=-\log(\tilde t)+\text{const.}$. The inset shows the time constant of final relaxation, where $I\propto \exp(-\ndt/\tau)$.}
\label{fig:mutual}		
\end{figure}

\begin{figure*}			
\includegraphics[width=\textwidth]{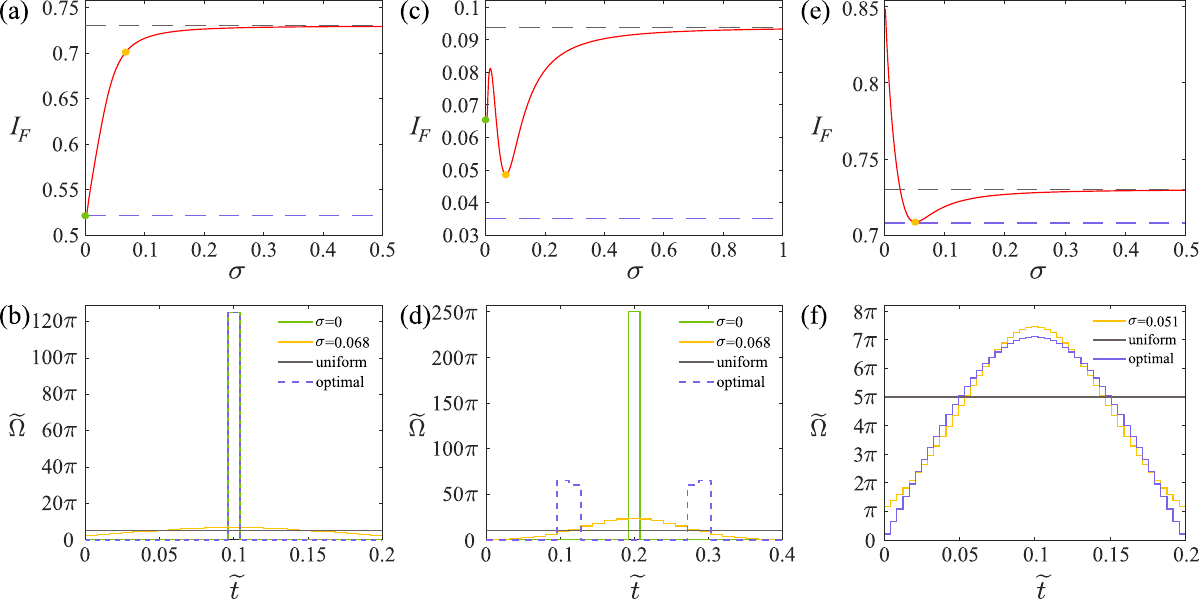}
 \caption{Efficiency of non-uniform mixing sequences, $\ndOME(\ndt)$. Panels (a) and (c) compare sequences with a fixed total rotation angle and panel (e) with a fixed dissipation. (a) Mixing efficiency of the rotation sequence with a Gaussian dependence $\ndOME \sim \exp[-(\ndt-\ndt_F/2)^2/2\sigma^2]$, discretized with 25 segments, for  $\int{\ndOME}d\ndt=\pi$, $\ndt_F=0.2$ (red line).  The blue dashed line shows the optimal and the gray dashed line the uniform sequence. (b) The sequences used in panel (a). (c) As in (a), but with $\int{\ndOME}d\ndt=4\pi$, $\ndt_F=0.4$. (d) The sequences used in panel (c). The optimal sequence becomes bimodal. (e) Mixing efficiency of Gaussian sequences, discretized with 49 segments, with equal total dissipation $\int{\ndOME^2}d\ndt=5\pi^2$ and $\ndt_F=0.2$. (f) The sequences used in (e).
}
\label{fig:6figure}		
\end{figure*}

The decay of mutual information at different constant rotational velocities is shown in Fig.~\ref{fig:mutual}. In all cases $I(t)$ monotonically decreases with time (as in any Markovian process, information can only be lost, but never recovered), but we see that the decay is accelerated by increasing the rotation rate, proving that the interplay between advection and diffusion accelerates the mixing process as quantified with the loss of mutual information.

A more conventional way of quantifying mixing consists of studying the dissolution of an initial pattern \cite{Danckwerts1952}. For example, the initial state can consist two different fluids.
A comparison between the decay of concentration variance for three such patterns and mutual information $I(t)$ is shown in Figs.~\ref{fig:pattern} and \ref{fig:var}. It shows that the slowest of the patterns has the same final decay rate as $I(t)$ while others are faster, in line with the argument that mutual information provides the strictest possible measure of mixing efficiency.

At short times, the effect of advection is small and the particle dynamics can be approximated as free diffusion. With a starting position $\mathbf{x}_0$ sufficiently far from the walls, the evolution of entropy can be approximated as $S[\mathbf{x}_t|\mathbf{x}_0] = \log{t}+\log{4\pi D} +1$ \cite{Tang.Golestanian2020}. Inserting the conditional entropy into Eq.~\eqref{3E} and switching to non-dimensional units gives an approximation for the mutual information $I(\ndt)= -\log{t}-\log{4\pi} -1 +\log(\tilde A)$, where $\tilde A$ is the dimensionless surface area. The result is shown by the dashed line in Fig.~\ref{fig:mutual}. While giving the correct slope, the value of $I$ is slightly underestimated, which we attribute to the effect of reflective boundaries.

Because mixing in viscous fluids requires an interplay between advection and diffusion, we expect that the mixing efficiency depends on the time sequence of rotation $\ndOME(\ndt)$ and not just the total angle $\int_0^{\ndt_F} \ndOME d\ndt$ \cite{Villermaux2019}. For example, rotation at the very beginning or the end of the time interval just reorders the positions, but does not reduce the mutual information. We also know, as shown above for any mixing process, that $I$ is invariant if we reverse the mixing sequence in time, $\Omega(t)\to \Omega(t_F-t)$.

We investigated the mixing efficiency of time-dependent advection under two different conditions: i) for a fixed total rotation $\int_0^{t_F}{\ndOME}d\ndt$ with the additional condition that the sense of rotation be constant ($\ndOME>0$) and ii) for a fixed total viscous dissipation, $\int_0^{t_F}{\ndOME^2}d\ndt$. In each of the settings, we compared the mixing efficiency of uniform rotation ($\ndOME=\rm const.$) with a Gaussian profile, centered around the middle of the time interval $\ndOME \propto \exp[-(\ndt-\ndt_F/2)^2/2\sigma^2]$, as well as the globally optimal sequence. All sequences were discretized with an odd number of time intervals (25 or 49). We determined the optimal sequence using a global optimizer [GlobalSearch in MATLAB (MathWorks Inc.)], using the velocities in individual intervals as optimization variables.  For a constant rotation, the dependence on the width of the Gaussian is usually monotonic [Fig.~\ref{fig:6figure}(a)] and optimal mixing is achieved by a sharp, discrete rotation at midtime [Fig.~\ref{fig:6figure}(b)]. For certain parameter ranges, especially for large total rotation angles and long times, the dependence becomes non-monotonic [Fig.~\ref{fig:6figure}(c)]. In such cases, the optimal sequence consists of two symmetrically arranged peaks [blue dashed line in Fig.~\ref{fig:6figure}(d)]. If the total viscous dissipation is kept constant, mixing sequences with strong non-uniformity naturally become less efficient [Fig.~\ref{fig:6figure}(e)]. Then the optimal velocity profile becomes approximately parabolic with a maximum in the middle and dropping to zero at the beginning and end of the interval. Overall, the analysis shows that the velocity profiles that maximize the mixing efficiency are always non-uniform, regardless whether the total rotation or the total viscous dissipation is kept constant. The symmetry of mixing efficiency upon time reversal is also reflected in the symmetry of the optimal profiles. The exact dependence, however, is complex and can be non-monotonic in certain parameter regions.

As an alternative approach, mutual information can also be estimated using lossless data compression algorithms. The (information) entropy of a dataset in principle gives a lower bound on its compressed size. By creating a file with initial and final positions of particles $(\mathbf{x}_0,\mathbf{x}_t)$ and compressing it, we obtain an upper bound on the joint entropy $S[\mathbf{x}_0,\mathbf{x}_t]$, and consequently, a lower bound on the mutual information $I(t)$ using Eq.~\eqref{2E}. As each position is a two-component vector, the problem is equivalent to computing entropy of a distribution in a 4-dimensional space, larger than in most uses of compression algorithms for entropy evaluation reported to date.

\begin{figure}								
\centering\includegraphics[width=0.8\columnwidth]{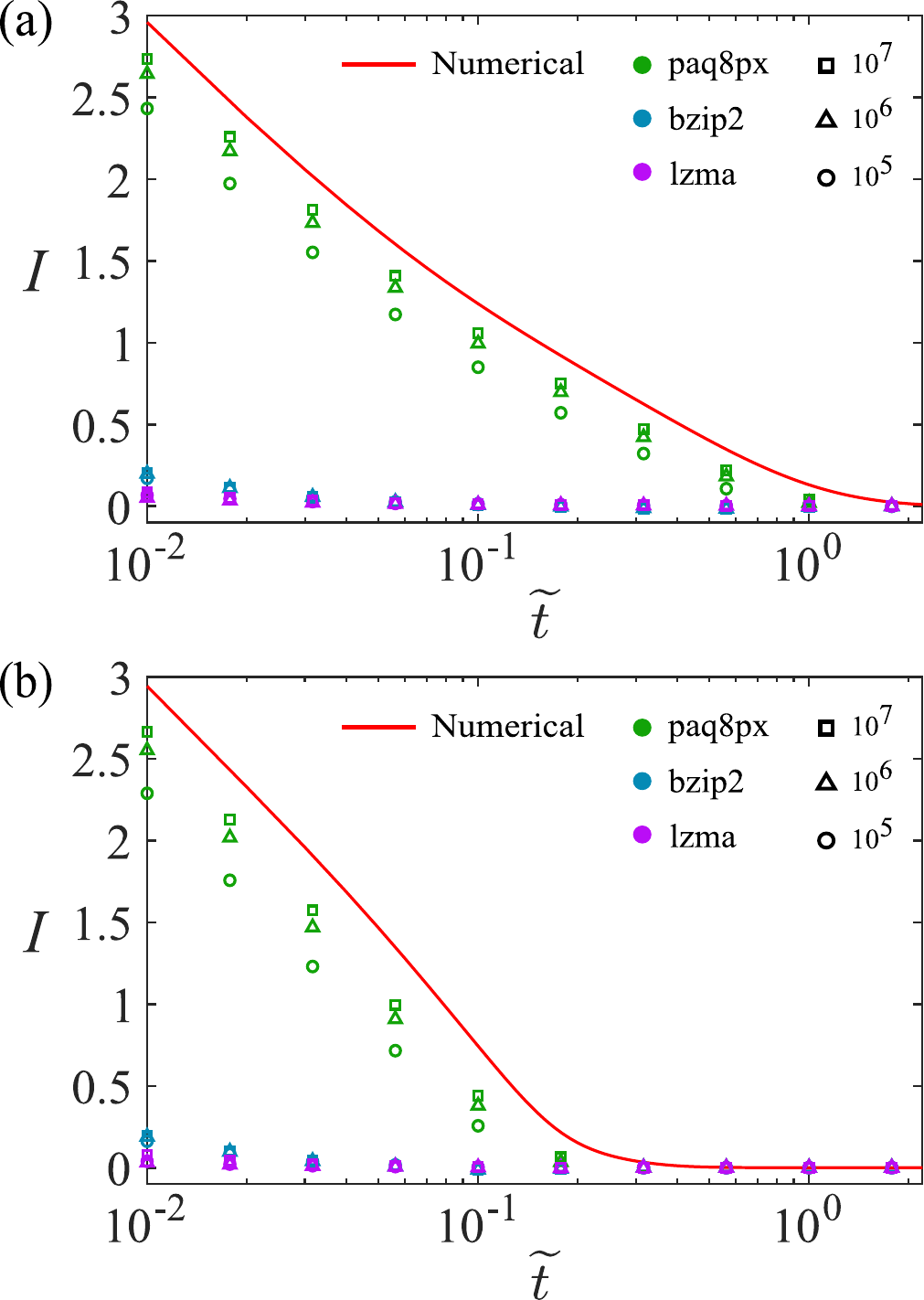} 
\caption{Mutual information between the initial state and the state after time $\ndt$, estimated by means of three different compression algorithms (color). Different symbol shapes mark different data set sizes $N$. The red line shows the numerical result using the spectral solution. (a) Process with pure diffusion $\ndOME=0$. (b) Process with both diffusion and advection $\ndOME=20\pi$.}
\label{fig:2D}		
\end{figure}

We generate the data set by randomly selecting an initial position $\mathbf{x}_0$ with uniform density inside the fluid area. For each initial position, we simulate the
Brownian trajectory with the Euler-Maruyama method where each next position is determined as
\begin{equation}
\mathbf{x}_{t+\Delta t} = \mathbf{R}\left[\omega\left(\left| \mathbf{x}_t\right|\right) \Delta t\right]\cdot \mathbf{x}_{t} + \boldsymbol{\xi}_{t}.
\end{equation}
Here $\mathbf{R}(\Delta \theta)$ denotes a rotation matrix with the rotation angle $\Delta \theta$ and $\boldsymbol{\xi}_t$ is a vector where each component is a Gaussian-distributed random variable with mean zero and standard deviation $\sqrt{2 D \Delta t}$. The combination of rotational motion and noise in Cartesian coordinates is chosen to avoid spurious drift caused by the integration procedure. Positions outside the radial range $[R_{\rm min},R_{\rm max}]$ were reflected at the boundary. Both positions were expressed in polar coordinates and the quantities $r_0^2-R_{\rm min}^2$, $\theta_0$, $r_F^2-R_{\rm min}^2$, $\theta_F$ were normalized, converted to 8-bit unsigned integers and written to a file in this order. Using $r^2$ as a variable ensures a homogeneous radial distribution. The simulation is repeated with $N=10^5$, $10^6$ or $10^7$ starting points, giving a file size of $400\,\rm kB$, $4\,\rm MB$ or $40\,\rm MB$, respectively. The optimally compressed size of this file ($C_t$) is a measure of the joint entropy $S[\mathbf{x}_0,\mathbf{x}_t]$ up to a constant that depends on the level of coarse graining. Likewise, the compressed size of an equivalent file with uncorrelated random initial and final positions ($C_\infty$) is a measure of the sum of the entropies of the initial and the final distributions $S[\mathbf{x}_{0}]+S[\mathbf{x}_{t}]$. Following Eq.~\eqref{2E}, the mutual information can be estimated as
\begin{equation}
     \label{7E}
     I[\mathbf{x}_{0};\mathbf{x}_{t}] \approx (\log{2}) \frac{C_\infty - C_t}{N}
\end{equation}
with both file sizes measured in bits.

For data compression, we use three different programs: the commonly used Lempel–Ziv–Markov chain algorithm (LZMA) and {\sc bzip2}, as well as the experimental, neural network based algorithm PAQ8PX \cite{paq8px,Mahoney2005,Knoll:2012}. A simple test case comprising a 1-dimensional drift-diffusion process shows that all three algorithms qualitatively give the right dependence when 10-bit integers are used, but PAQ8PX consistently gave errors $\lesssim 0.1$ with large samples (Fig.~\ref{fig:1D}).  Figure \ref{fig:2D} shows the mutual information $I(\ndt)$ of our mixing process estimated using the three compression algorithms and different dataset sizes in comparison with the numerical results, obtained with the spectral solution. The comparison is shown for a diffusive process [Fig.~\ref{fig:2D}(a)] and for constant advection with $\ndOME=20\pi$ [Fig.~\ref{fig:2D}(b)]. Whereas the standard algorithms practically failed to detect any mutual information, PAQ8PX consistently gave results with an absolute error $\ll 1$. With the largest size $N$, the errors are consistently $\lesssim 0.3$.

In conclusion, we have introduced mutual information between the initial and the final state as a universal measure for mixing efficiency in microfluidic setups with strong interplay between advection and diffusion. We have shown that under this measure, the mixing efficiency is symmetric upon time reversal of the actuation sequence. Among all sequences with the same rotation angle, the ones with optimal mixing consist of a fast rotation in the middle of the time interval, or in some cases two symmetrically arranged. We have also demonstrated that advanced neural network based compression algorithms can be applied to estimate mutual information to a high accuracy. The latter can prove useful in more complex flows in which a full solution of the advection-diffusion equation may not be tractable. We also stress that the Couette flow that we chose as a demonstration is far from optimal and that several works have investigated sequences with optimal kinematics under given mixing norms \cite{Eggl.Schmid2020, Eggl.Schmid2022, GUBANOV.CORTELEZZI2010, LIN.DOERING2011}. Finding the mixing pattern without geometric restrictions that minimizes mutual information remains a task for future. Furthermore, we expect that our formalism will also be applicable to more complex mixing situations, for example by active swimmers \cite{Saintillan.Shelley2008,Mueller.Thiffeault2017,Reinken.Wilczek2022}, natural or artificial cilia \cite{Toonder.Anderson2008,Shields.Superfine2010,Rahbar.Gray2014,Ding.Kanso2014, Jonas.Choma2013, Nawroth.McFallNgai2017} or in active materials.

\acknowledgments
We thank Alja\v{z} Godec and Reiner Kree for helpful discussions.
This work has received support from the Max Planck School Matter to Life and the MaxSynBio Consortium, which are jointly funded by the Federal Ministry of Education and Research (BMBF) of Germany, and the Max Planck Society. AV acknowledges support from the Slovenian Research Agency (Grant No.\ P1-0099).

\bibliography{ref}

\clearpage
\appendix
\renewcommand{\thefigure}{A\arabic{figure}}
\setcounter{figure}{0}
\begin{widetext}
\section{1D test case for mutual information estimation using compression algorithm}

As a test case for mutual information estimation using compression algorithm, we use a drift-diffusion process on a 1D ring. The initial positions of particles on the ring are uniformly distributed. The final position of the same particle is 
determined as $\phi_0+\Delta \phi$, where $\Delta \phi$ is Gaussian distributed with mean $\mu$ and variance $\sigma$. The mutual information between initial and final position can be calculated (see Eq.\ 2 in the main text) as 
\begin{equation}
\label{1S}
     I[\phi_{0};\phi_{t}] = S[\phi_{t}]-S[\phi_{t}|\phi_{0}]= S[\phi_{t}]+\int_0^{2\pi}{P(\phi_0)\int_0^{2\pi}{P(\phi_t|\phi_0)\log P(\phi_t|\phi_0)}d \phi_t}d \phi_0
\end{equation}
where $S[\phi_{t}] = \log{2\pi}$, $P(\phi_0) ={1}/(2\pi)$ and 
\begin{equation}
\label{2S}
P(\phi_t|\phi_0) = \sum_{k=-\infty}^{\infty} \frac{1}{\sqrt{2\pi}\sigma}\exp\left[{-\frac{(\phi_t+2\pi k-\mu-\phi_0)^2}{2\sigma^2}}\right] \notag
= \frac{1}{2\pi}\vartheta_3\left(\frac12 (\phi_t-\mu-\phi_0), e^{-\sigma^2/2}\right).
\end{equation}
Here, $\vartheta_3(z, q)=\sum_{k=-\infty}^{\infty}q^{k^2}e^{2kiz}$ is a Jacobi theta function.

To test the compression algorithms,  we generate a set of $N=10^5$, $10^6$ or $10^7$ pairs of initial and final positions $(\phi_0,\phi_t)$, converted them to $10$ or $16$ bit unsigned integers and wrote them to a file. Each integer was stored as 2 bytes regardless of the number of non-zero bits. The compressed file size, $C_t$ is then compared with the size of an equivalent file with uncorrelated initial and final distributions, $C_\infty$. 
Based on main text Eq.\ 2, we estimate the mutual information as $I_F[\phi_{0};\phi_{t}]=(\log 2) (C_\infty-C_t)/N$.
The result of three lossless compression algorithms: bzip2, Lempel–Ziv–Markov chain algorithm (LZMA) and PAQ8PX is compared with the theoretical result in Fig.~\ref{fig:1D}.

\begin{figure}[h!]			
\includegraphics[width = 16.5cm]{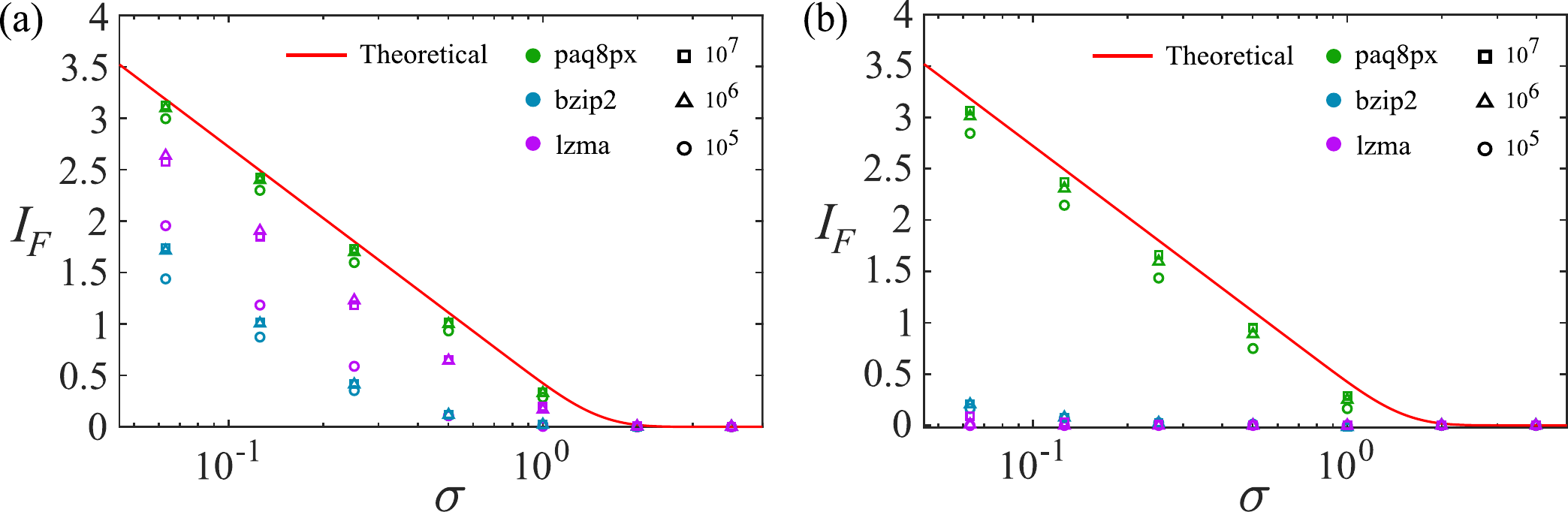}
\caption{Theoretical result for the mutual information of a 1D drift-diffusion process as a function of the distribution variance $\sigma$ compared with the estimates using 3 compression algorithms (color) and 3 data set sizes $N$ (symbol type). The data are stored as 10-bit integers in (a) and 16-bit integers in (b).}
\label{fig:1D}		
\end{figure}

\section{Comparison between mutual information and variance of concentration}
Mutual information is a measure of mixing efficiency which does not depend on a specific initial distribution of the solute. 
Alternatively, if the initial distribution of two fluids is known, the variance in their concentration, which is equivalent to the probability distribution $P$ used in the main text, can be used to quantify mixing \cite{Danckwerts1952}. 
In the following, we make a comparison between the two measures for different initial patterns. We initialize the concentration to $c=1$ in half of the volume, distributed over 1, 2 or 3 segments (Fig.~\ref{fig:pattern}).
We then apply the same spectral method as in the main text and numerically solve the advection-diffusion equation 
$\partial_{t} c + \boldsymbol{\nabla}\cdot(\mathbf{v}c)=D\Delta c$.
The resulting concentration distribution over time is shown in Fig.~\ref{fig:pattern}. The  variance of concentration is defined as $\sigma_{c}^2=\langle( c - \langle c\rangle)^2\rangle$, where the brackets $\langle \rangle$ indicate spatial averages and $\langle c \rangle=1/2$ by the choice of initialization. Figure~\ref{fig:var} shows a comparison between the variance for different patterns and mutual information $I_F$ with pure diffusion (a) and with uniform advection (b).  In the long time limit, mutual information has the same exponential decay rate as the variance of pattern 1, which has the slowest mixing. The variance of higher two patterns decays faster than mutual information. 
This comparison is in line with the theoretical argument that mutual information provides the strictest possible measure of mixing efficiency.

\begin{figure}								
\includegraphics[width = 15.6cm]{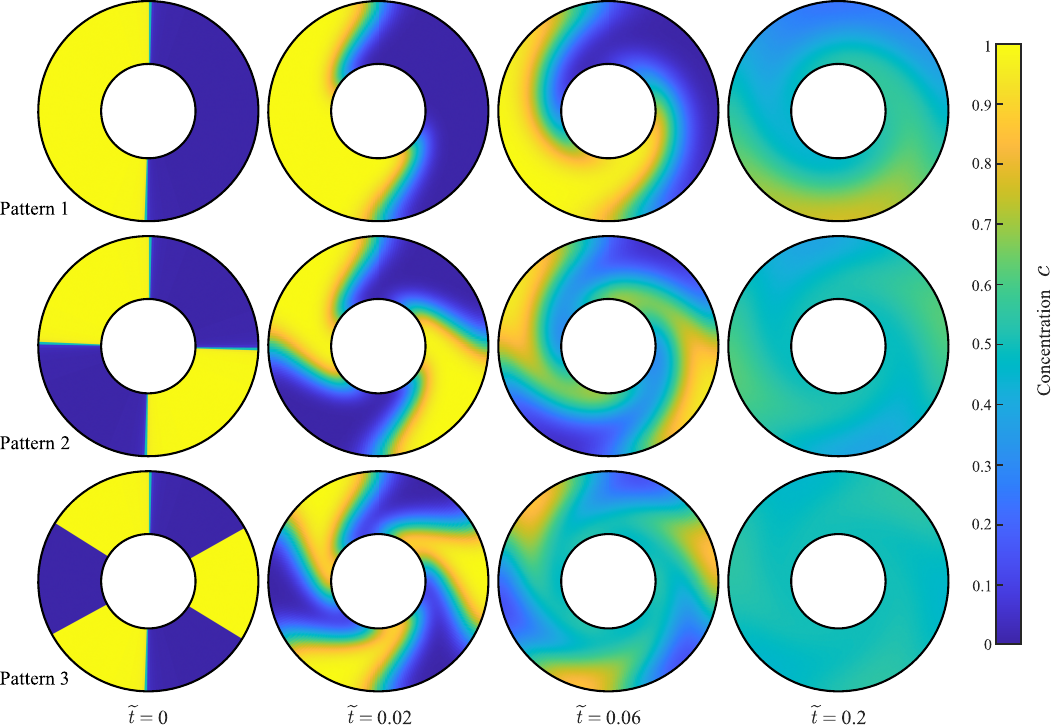}
\caption{Time evolution of the concentration $c(\mathbf{x},\ndt)$, with three different initial patterns. In pattern 1, 2, 3, the whole unit is divided into 2, 4, 6 equal parts, respectively. All the processes involve both diffusion and uniform advection ($\ndOME=20\pi$).
}
\label{fig:pattern}		
\end{figure}

\begin{figure}								
\includegraphics[width = 17.5cm]{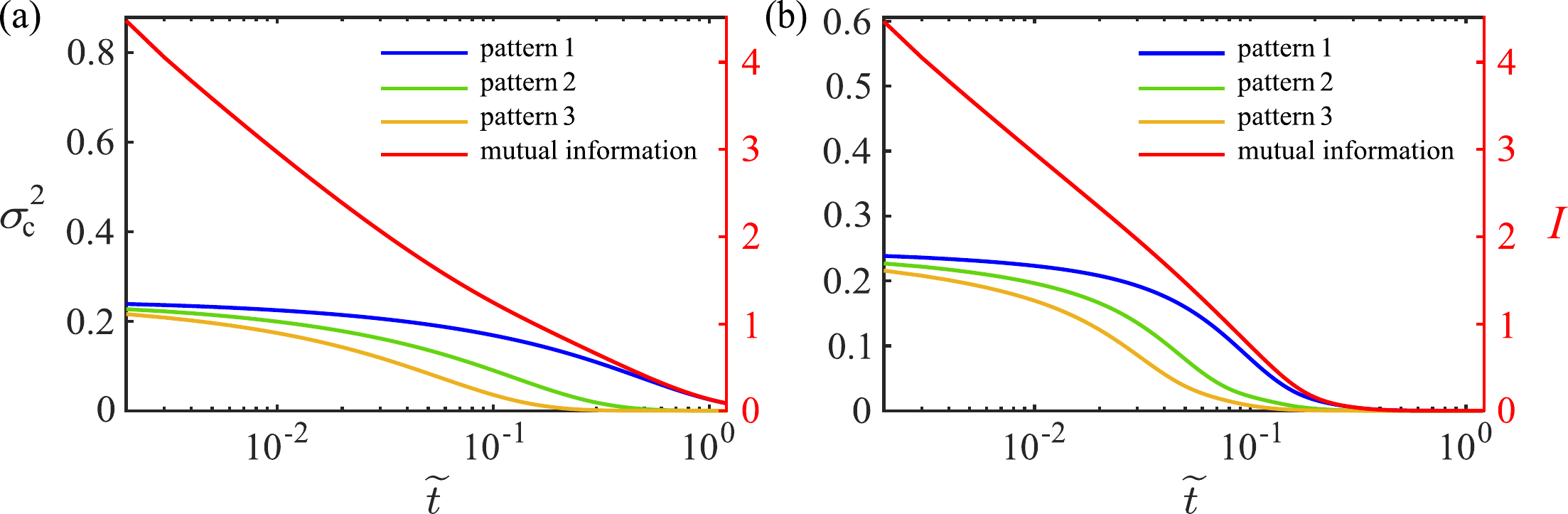}
\caption{Numerical results for the variance of concentration with different initial patterns (color; left axis) compared with the result of the mutual information (red; right axis). (a) Processes with pure diffusion ($\ndOME=0$). (b) Processes with both diffusion and advection ($\ndOME=20\pi$).}
\label{fig:var}		
\end{figure}

\end{widetext}

\end{document}